\documentclass[submission,copyright,creativecommons]{eptcs}
\providecommand{\event}{QPL 2023} 
\usepackage{iftex, amssymb, amsmath, amsfonts, tikz-cd, mathtools}

\ifpdf
\usepackage{underscore}         
  \usepackage[T1]{fontenc}        
\else
  \usepackage{breakurl}           
\fi

\title{Locally Tomographic Shadows (Extended Abstract)}

\author
{
Howard Barnum
\email{hnbarnum@aol.com}
\and
Matthew A.\ Graydon  
\author{Matthew A.\ Graydon}
\institute{Institute for Quantum Computing\\ University of Waterloo}
\email{\quad m3graydo@waterloo.ca}
\and 
Alex Wilce
\institute{
Susquehanna University}
\email{\quad wilce@susqu.edu}
}
\def\titlerunning{Locally Tomographic Shadows}
\def\authorrunning{Barnum, Graydon, Wilce}

\newcommand{\Vstar}[1]{\V^{\ast}{#1}}
\renewcommand{\H}{{\mathbf H}}
\newcommand{\J}{{\mathbb J}}
\newcommand{\K}{{\mathbf K}}
\renewcommand{\L}{{\mathcal L}}

\newcommand{\E}{{\mathbb V}^{\ast}} 
\newcommand{\V}{{\mathbb V}}
\newcommand{\Vst}{{\mathbb V}^{\ast}}

\renewcommand{\prod}{\bullet}

\newcommand{\Cat}{{\mathcal C}}
\newcommand{\CPM}{\mbox{\bf CPM}}
\newcommand{\loCat}{\mbox{Loc}({\mathcal C}, \V)} 

\newcommand{\C}{{\mathbb C}}

\renewcommand{\ker}{\mbox{Ker}}
\newcommand{\Ran}{\mbox{Ran}}
\newcommand{\R}{{\mathbb R}}
\newcommand{\LT}{\mbox{LT}}

\newcommand{\InvQM}{\mbox{\bf InvQM}}
\newcommand{\CQM}{\mbox{${\mathbb C}$QM}}

\newcommand{\CP}{\mbox{ CP}}

\newcommand{\RQM}{\mbox{$\boldsymbol \R${\bf QM}}}
\newcommand{\RHilb}{\mbox{$\boldsymbol \R${\bf Hilb}}}

\newcommand{\Prob}{\mbox{\bf Prob}}

\newcommand{\Blambda}{\tilde{\V}}

\renewcommand{\tilde}{\widetilde}

\newcommand{\bigmaxtensor}{\bigotimes_{\text{max}}}
\newcommand{\bigmintensor}{\bigotimes_{\text{min}}}

\newcommand{\maxtensor}{\otimes_{\mbox{max}}}
\newcommand{\mintensor}{\otimes_{\mbox{min}}} 

\newcommand{\tempout}[1]{{}}
\newcommand{\Tr}{\mbox{Tr}}
\newcommand{\tr}{\Tr}
\newcommand{\1}{{\mathbf 1}}

\newcommand{\Sym}{\mbox{Sym}}

\newcommand{\ltensor}{\boxtimes}

\newcounter{thaler}
\newenvironment{mlist}{\begin{list}{\arabic{thaler}}%
{\usecounter{thaler}
\setlength{\rightmargin}{\leftmargin}
\topsep=0pt
\itemsep=0pt
\parskip=0pt
\parsep=0pt
}}{\end{list}}

\newcommand\reallywidecheck[1]{%
\savestack{\tmpbox}{\stretchto{%
\scaleto{%
\scalerel*[\widthof{\ensuremath{#1}}]{\kern-.6pt\bigvee\kern-.6pt}%
{\rule[-\textheight/2]{1ex}{\textheight}}
}{\textheight}%
}{0.5ex}}%
\stackon[1pt]{#1}{\tmpbox}%
}

\begin{document}
\maketitle



\begin{abstract}
Given a monoidal probabilistic theory --- a symmetric monoidal category $\cal C$ of systems and processes, together with a functor $\bf V$ assigning concrete probabilistic models 
to objects of $\Cat$ --- 
we construct a locally tomographic probabilistic theory $\LT(\Cat, \V)$ --- the {\em locally tomographic shadow} of $(\Cat, \V)$ ---  describing  phenomena observable by local agents controlling systems in $\Cat$, and able to pool information about joint measurements made on those systems. Some globally distinct states become locally indistinguishable in $\LT(\Cat, \V)$, and we  restrict the set of processes to those that respect this 
indistinguishability. This construction is investigated in some detail for real quantum theory. 
\end{abstract}

\section{Introduction}

As is well known, complex quantum theory is distinguished from its real counterpart by a property 
called {\em tomographic locality}
 or {\em local tomography} \cite{BW12, Hardy}: states of a bipartite system are entirely determined by the joint probabilities they assign to measurements performed 
separately on the two component systems. That this fails for finite-dimensional real quantum theory is evident on dimensional grounds, but 
it's more instructive to note that 
if $\H$ and $\K$ are real Hilbert spaces, 
the space $\L_{s}(\H \otimes \K)$ of bounded self-adjoint operators on $\H \otimes \K$ 
decomposes as $\L_{ss} \oplus \L_{aa}$, where 
$\L_{ss} := \L_{s}(\H) \otimes \L_{s}(\K)$ and 
$\L_{aa} := \L_{a}(\H) \otimes \L_{a}(\K)$, with $L_{a}(\H)$ the space of bounded 
anti-self adjoint operators on $\H$, and similarly for $\K$. This is 
an orthogonal decomposition with respect to the trace inner product. Thus, 
a density operator $\rho$  on $\H \otimes \K$ has a unique decomposition 
$\rho_{ss} + \rho_{aa}$, with $\rho_{ss} \in \L_{ss}$ and $\rho_{aa} \in \L_{aa}$. 
Given effects $a \in \L_{s}(\H)$ and $b \in \L_{s}(\K)$, $a \otimes b$ lives in $\L_{ss}$, 
so $\Tr((a \otimes b)\rho_{aa}) = 0$, and thus $\Tr((a \otimes b)\rho) = \Tr((a \otimes b)\rho_{ss})$.  In other words, states with the same $\L_{ss}$ component but distinct 
$\L_{aa}$ components are {\em locally indistinguishable}.\footnote{In contrast, 
in CQM, any anti-selfadjoint operator $a$ has the form $ib$ where $b$ is self-adjoint. 
Hence, if $a = ib$ and $a' = i b'$ are anti-self adjoint, $a \otimes a'$ (as an element of $\L_{s}(\H \otimes \K)$) coincides with $-(b \otimes b') \in \L_{s}(\H) \otimes \L_{s}(\K)$. 
That is: $\L_{ss} = \L_{aa} = \L_{s}(\H \otimes \K)$.}

This suggests a construction: simply ``factor out" the non-locally observable $\L_{aa}$ degrees of freedom to obtain a locally tomographic theory. Here, we not only show that this is feasible, but go much further. 
By a {\em monoidal probabilistic theory} we mean a pair $(\Cat, \V)$ 
where $\Cat$ is a symmetric monoidal category and $\V : \Cat \rightarrow \Prob$ is a functor from a symmetric monoidal category $\Cat$ to a suitable category 
of concrete probabilistic models, taking monoidal products in $\Cat$ 
to (non-signaling) composites in $\Prob$.  We outline this framework in Section 2.\footnote{Because this slightly extends the standard 
GPT framework as described, e.g., in \cite{BWFoils}, we go into a bit 
of detail here} Given a such a theory $(\Cat, \V)$, 
we construct a new, locally tomographic theory, $\LT(\Cat,\V)$, that describes what the world ``looks like" to local agents, at least if one restricts attention to processes 
that respect local indistinguishability. Taking a cue from Plato, we call this 
the {\em locally tomographic shadow} of the original theory.  This is described in 
Section 3. In Section 4, we return to the case of real QM, and show that our construction 
leads to a non-trivial theory that differs from both real and complex QM, but still allows for 
entangled states and effects.  

If one lifts the restriction on processes mentioned above, it is still possible to construct 
a ``shadow" theory, at the price of accepting an additional layer of non-determinism 
on top of that built into the probabilistic structure of the original model. This is briefly discussed in the concluding section 5, along with a number of other directions for further work. 

This is a preliminary sketch of a longer work in progress. 

\section{Generalized Probabilistic Theories}

For our purposes, a {\em probabilistic model} \cite{BWInf, BWFoils} is pair $(\V,u)$ where $\V$ is an ordered 
real vector space and $u$ is a strictly positive linear functional thereon, referred to as the {\em unit effect} of the model.  Elements 
$\alpha$ of $\V_+$ with $u(\alpha) = 1$ are the 
(normalized) {\em states} of the system. {\em Effects} (measurement 
results) are positive functionals $a$ on 
$\V$ with $a \leq u$; $a(\alpha)$ is the probability of $a$'s of occurring in state $\alpha$.  Where we wish to 
keep track of several models, we label them $A$, $B$, 
etc., writing, e.g., $\V(A)$, $\Vstar(A)$ for the 
associated ordered vector spaces and their 
duals, and $u_{A}$, for the unit effects. 
A {\em process} from a probabilistic model $(\V(A),u_A)$ to a probabilistic 
model $(\V(B),u_B)$ is a positive linear mapping $\phi : \V(A) \rightarrow \V(B)$ with $u_{B}(\phi(\alpha)) \leq u_{A}(\alpha)$ for every $\alpha \in \V(A)_+$ 

We write $\Prob$ for the category of probabilistic models and processes.  
In the broadest sense, a probabilistic theory is simply a functor $\V$ from some category $\Cat$ into $\Prob$.\footnote{Note that such a functor $\V$ comes with a designated unit functional $u_{A} \in \V(A)^{\ast}$ for every object $A \in \Cat$ with 
$u_{A} \circ \V(\phi) \leq u_{B}$ 
for all $\phi \in \Cat(A,B)$.} $\Cat$ 
can be understood to consist of ``actual" physical systems and 
processes, or of any mathematical proxies for these (classical 
phase spaces, Hilbert spaces, open regions of spacetime, spin networks, 
or what-have-you). $\V$ attaches probabilistic 
apparatus to these systems and processes in such a way as, e.g., to describe the possible experiences of agents. 

In what follows, we denote such a probabilistic theory as a pair $(\Cat, \V)$. 
We assume that $\V$ is {\em injective on objects}, which makes $\V(\Cat)$ a subcategory 
of $\Prob$.  This assumption holds for all of the examples we wish to 
consider.  Where $\V$ is also injective on morphisms, 
we say that $(\Cat, \V)$ is {\em process tomographic} \cite{Chiribella}, in which 
case, we can simply identify $\Cat$ with the corresponding subcategory of $\Prob$.  Note that 
the latter --- more exactly, the probabilistic theory  
$(\V(\Cat),I)$, where $I : \V(\Cat) \rightarrow \Prob$ is the inclusion functor --- is automatically 
process tomographic, regardless of whether or not $(\Cat, \V)$ is so.

The example of primary interest here takes $\Cat = \CPM_{\R}$, the category of finite-dimensional real Hilbert spaces with morphisms $\H \rightarrow \K$ given by 
completely positive linear mappings $\L(\H) \rightarrow \L(\K)$. In contrast to the complex case, there exist linear mappings $\L_{s}(\H) \rightarrow \L_{s}(\K)$ that preserve positivity but not adjoints \cite{CDPR}. We reserve the term positive for those linear mappings that preserve both. Equivalently, $\phi : \L(\H) \rightarrow \L(\K)$ is positive iff $\phi$ maps positive operators to positive operators, hence mapping $\L_{s}(\H)$ to $\L_{s}(\K)$, and also maps $\L_{a}(\H)$ to $\L_{a}(\K)$. The probabilistic theory we have in mind when we speak of real QM is then $(\CPM_{\R}, \V)$ where $\V(\H) = \L_{s}(\H)$, the latter understood as an order unit space with $u_{\H} = \Tr( \cdot )$.  In contrast to the complex case, the restriction of 
$\phi$ to the self-adjoint part, $\L_{s}(\H)$, 
of $\L(\H)$ does not determine $\phi$, so $\RQM$ is not process tomographic. 

{\bf Composite Models} A (non-signaling) {\em composite} of models $\V(A)$ and $\V(B)$ is a model $\V(AB)$, together with a pair of bilinear mappings
\[m : \V(A) \times \V(B) \rightarrow \V(AB) \ \mbox{and} \ \  \pi : \V(A)^{\ast} \times \V(B)^{\ast} \rightarrow \V(AB)^{\ast}\]
such that (i) $\omega \circ \pi$ is a joint state on $A$ and $B$ for all $\omega \in \V(AB)$, and (ii) $\pi(a,b)m(\alpha, \beta) = a(\alpha)b(\beta)$ for all $a \in \Vst(A), b \in \Vst(B)$ and all $\alpha \in \V(A)$, $\beta \in \V(B)$. 

We think of $\pi(a,b)$ as representing the joint outcome $(a,b)$ of a pair of experiments performed on $A$ and $B$, and $m(\alpha, \beta)$ as the result of preparing states $\alpha$ and $\beta$ independently on $A$ and $B$. We refer to $\pi(a,b)$ and $m(\alpha, \beta)$ as {\em product} effects 
and states, respectively. 

If $\H_{A}$ and $\H_{B}$ are two complex (finite-dimensional) Hilbert spaces, the obvious choice for a composite model is $\V(AB) = \V(\H_{A} \otimes \H_{B}) 
= \L_{s}(\H_{A} \otimes \H_{B})$, with $\pi(a,b)(\omega) = \Tr((a \otimes b)\omega)$ and $m(\alpha, \beta) = \alpha \otimes \beta$. 

If every state $\omega \in \V(AB)$ is determined by the joint 
probabilities $\omega(\pi(a,b))$, we say that $AB$ is {\em locally tomographic}. 
Given arbitrary models $\V(A)$, $\V(B)$, there are two 
extremal locally tomographic composites, obtained by endowing $\V(A) \otimes \V(B)$ with 
the {\em minimal} and {\em maximal} tensor cones \cite{BWFoils, Wilce92, Wittstock}. 
The former is generated by {\em separable} states, i.e., convex combinations of product states. The latter consists of all bilinear forms that are 
positive on products of effects. 
We write $\V(A) \otimes_{\mbox{min}} \V(B)$ and $\V(A) \otimes_{\mbox{max}} \V(B)$ 
for $\V(A) \otimes \V(B)$ as ordered by these minimal and maximal cones, respectively. A composite 
$\V(AB)$ is locally tomographic iff $m : \V(A) \otimes \V(B) \rightarrow \V(AB)$ is a linear isomorphism, and in this case it's usual simply to 
identify the two spaces. One then has 
\[(\V(A) \mintensor \V(B))_{+} 
\leq \V(AB)_{+} \leq (\V(A) \otimes_{\mbox{max}} \V(B))_{+}.\] 
If $A$ and $B$ are quantum models, the inclusions are proper: 
$(\V(A) \mintensor \V(B))_+$ contains only separable states, while $(\V(A) \maxtensor \V(B))_+$ 
contains states corresponding to non-positive operators \cite{KRF, Wilce92}.  
More generally, $\V(A)\mintensor\V(B)$ allows only separable states, but 
arbitrarily entangled effects; the maximal 
tensor product $\V(A) \otimes_{\mbox{max}} \V(B)$ allows the reverse. Both tensor products are naturally associative, and extending straightforwardly to tensor products of more than 
two factors.

{\bf Definition:} A {\em monoidal probabilistic theory} is a structure $(\Cat, \V, m, \pi)$ 
where $\Cat$ is a symmetric monoidal category, $\V$ is a functor $\Cat \rightarrow \Prob$, 
and $m$ and $\pi$ are assignments, to every pair of objects $A, B \in \Cat$, of 
bilinear mappings 
\[m_{A,B} : \V(A) \times \V(B) \rightarrow \V(AB) \ \mbox{and} \ \pi_{A,B} : \Vst(A) \times \Vst(B) \rightarrow \Vst(AB)\footnote{We will 
write monoidal products in $\Cat$ juxtapositively, reserving the 
symbol $\otimes$ for the tensor product of linear spaces.}\]
such that 
\begin{mlist} 
\item[(i)] $m_{A,B}, \pi_{A,B}$ make $\V(AB)$ a composite of $\V(A)$ and $\V(B)$, 
\item[(ii)] $\V(I) = \R$ with 
\[m_{I,A} : \V(I) \times \V(A) \rightarrow \V(A), \ \pi_{I,A} : \Vst(I) \times \Vst(A) \rightarrow \Vst(A)\]
 the bilinear mappings uniquely defined by $m(1,\alpha) = \alpha$ and $\pi_{I,A}(1,a) = a$, and similarly 
for $m_{I,A}$ and $\pi_{A,I}$; 
\item[(iii)] $\V(\sigma_{A,B}) \circ \pi_{A,B} = \pi_{B,A}$, and similarly 
$m_{A,B} \circ \V(\sigma_{A,B}) = m_{B,A}$; and 
\item[(iv)] 
 for all morphisms $\phi : A \rightarrow A'$ and $\psi : B \rightarrow B'$ in $\Cat$, the diagram 
\begin{equation}
\begin{tikzcd}[row sep=huge]
\V(A) \otimes \V(B)  \arrow[dd, swap, "\V(\phi) \otimes \V(\psi)"] &  & \V(AB) \arrow[ll,swap,"\pi_{A,B}^{\ast}"] \arrow[dd, "\V(\phi \otimes \psi)"]\\
& & \\
 \V(A') \otimes \V(B')  & &
\V(A'B')  \arrow[ll,"\pi_{A',B'}^{\ast}"] 
\end{tikzcd}
\end{equation}
commutes. 
\end{mlist} 
If $\alpha \in \Cat(I,A)$, then by condition (ii), 
$\V(\alpha) : \R \rightarrow \V(A)$ defines an element of $\V(A)$, namely 
$\V(\alpha)(1)$. We make it a standing assumption in what follows that 
every normalized state in $\V(A)$ arises in this way.

{\em Remark:} 
On the left side of (1), $\V(\phi) \otimes \V(\psi)$ is the usual tensor product of 
linear maps, while on the right, $\phi \otimes \psi$ is the monoidal composite of the morphisms 
$\phi$ and $\psi$ in $\Cat$.  An equivalent statement is that, for all states 
$\omega \in \V(AB)$ and all effects $a' \in \Vst(A')$ and $b' \in \Vst(B')$, we have 
\[\omega(\V(\phi)^{\ast}(a'), \V(\psi)^{\ast}(b')) = \V(\phi \otimes \psi)(\omega)(a',b').\]
More compactly: $m$ and $\pi^{\ast}$ are natural transformations from the functor $\V \mintensor \V$ to the functor $\V \circ \otimes$, 
and from $\V \circ \otimes$ to $\V \otimes_{\mbox{max}} \V$, respectively. 
When no confusion seems likely, we will henceforth write $(\Cat, \V)$ for a monoidal probabilistic theory $(\Cat, \V, m, \pi)$. 

Evidently, $\RQM$, with its standard compositional structure, is an example of such a monoidal probabilistic theory.

All of the foregoing applies to composites of more than two models. One can show that 
both $\mintensor$ and $\maxtensor$ are both naturally associative.
 If one has an $n$-partite composite $A = A_{1} \cdots A_{n}$, built up  
recursively so that $A = (A_{1} \cdots A_{n-1})A_{n}$, then arguing inductively 
one has 
canonical positive mappings $\bigmintensor \V(A_{i}) \stackrel{m}{\longrightarrow} 
\V(A) \stackrel{\pi^{\ast}}{\longrightarrow} \bigmaxtensor \V(A_{i})$.

We call a monoidal probabilistic theory $(\Cat, \V)$ 
{\em locally tomographic} iff, for every pair of objects $A, B \in \Cat$, 
the composite $\V(AB)$ is locally tomographic. 

\tempout{
If $\V$ is injective on objects,  as we are assuming here, it gives rise naturally to a process-tomographic theory: the image category
$\V(\Cat)$ (more precisely,  the theory $(\V(\Cat), I)$, where $I : \V(\Cat) \rightarrow \Prob$ is the inclusion 
functor) is process tomographic
(as $I$ is injective on morphisms) whether or not $(\Cat, \V(\C))$ is so. 
}

As pointed out above, as long as $\V$ is injective on objects, as we are assuming here, process-tomography can be enforced by passing to the probabilistic theory. 
$(\V(\Cat), I)$ 
For local tomography, the situation is not so simple. Still, as we show in the next section, it is possible to construct, from a given probabilistic theory, a kind of locally tomographic quotient theory.

\section{Locally tomographic Shadows}

If a monoidal probabilistic theory $(\Cat, \V)$ is not locally tomographic, one can still ask what 
the world it describes would ``look like" to agents having access only 
to local measurements. As a first step, we need to assume that objects in $\Cat$ can be carved up in a preferred way into local pieces. To ensure this, we replace $\Cat$ with the  
its strictification, the category $\Cat^{\ast}$ having as objects, finite lists 
$\vec{A} = (A_1,...,A_n)$ of non-trivial (non-identity) objects
$A_i \in \Cat$, with the understanding that this represents the composite 
$\Pi_{i=1}^{n} A_i$ in $\Cat$, but with the indicated monoidal decomposition. Morphisms from $\vec{A} = (A_1,...,A_n)$ to $\vec{B} = (B_1,...,B_k)$ are simply 
be morphisms $\Pi_{i=1}^{n} A_i \rightarrow \Pi_{i=1}^{k} B_i$ in $\Cat$. 
This is a strict symmetric monoidal category, 
with $(A_1,...,A_n)(B_1,...,B_k) = (A_1,...,A_n, B_1,...,B_k)$,
 the empty sequence $( \, )$ serving as the monoidal unit. 
There is a (strong, but not strict) monoidal functor $\Pi : \Cat^{\ast} \rightarrow \Cat$ taking $\vec{A} = (A_1,...,A_n)$ to $\Pi \vec{A} := \Pi_{i=1}^{n} A_i$, with $\Pi ( \, ) = I_{\Cat}$, and acting as the identity on morphisms. This is not generally injective on objects (for instance, $\Pi(A,BC) = \Pi(A,B,C)$). 
Composing $\Pi$ with $\V$ now gives us a probabilistic theory $(\Cat^{\ast}, \V \circ \Pi)$ 
in which we have the desired canonical decompositions, but in which we've lost 
the desirable injectivity-on-objects feature. This feature will be recovered when we pass 
to the locally tomographic shadow, which we will now construct. 

\subsection{The shadow of a composite} 

If $AB$ is a composite of probabilistic models $A$ and $B$, 
then 
a state $\omega \in \Omega(AB)$ of the composite system determines a joint 
state  $\tilde{\omega} : \V(A)^{\ast} \times \V(B)^{\ast} \rightarrow [0,1]$, given 
by $\tilde{\omega}(a,b) := \pi_{A,B}(a,b)(\omega)$. This describes 
the joint probabilities of measurement outcomes carried out on $A$ and $B$ 
separately.  
We may call $\tilde{\omega}$ the {\em local shadow} of 
$\omega$. More generally, suppose that $\vec{A} := (A_1,...,A_n)$ is a sequence in $\Cat^{\ast}$ with 
composite  $\Pi \vec{A} := A$ in $\Cat$: as we've seen, there is a positive linear mapping 
\[\pi^{\ast}_{\vec{A}} : \V(A) \rightarrow \L^{n}(\Vst(A_1),...,\Vst(A_n))\]
taking $\omega \in \Omega(A)$ to the corresponding joint state $\tilde{\omega}$ 
on $A_1,...,A_n$ --- its local shadow --- given by 
\[\tilde{\omega}(a_1,...,a_n) := \pi^{\ast}_{A}(\omega)(a_1,...,a_n) = (a_1 \otimes \cdots \otimes a_n)(\omega)\]
for all $(a_1,...,a_n) \in \Vst(A_1) \times \cdots \times \Vst(A_n)$.

{\bf Definition:} For $\vec{A} = (A_1,...,A_n) \in \Cat^{\ast}$, let $\tilde{\V}(\vec{A})$ be the space $\bigotimes_{i} \V(A_i)$, 
ordered by the cone $\pi^{\ast}_{\vec{A}}(\V(\Pi \vec{A})_{+})$ 
consisting of local shadows of elements of $\V(\Pi \vec{A})_{+}$, 
and let $\tilde{u}_{A} = u_{A_1} \otimes \cdots \otimes u_{A_{n}}$. 
We call the probabilistic model $(\tilde{\V}(\Pi \vec{A}), \tilde{u}_{\vec{A}})$ 
the {\em locally tomographic shadow} of $A = \Pi\vec{A}$ with respect to the given 
decomposition (that is, with respect to $\vec{A}$). 

Going forward, it will be convenient to use the abbreviations 
$\LT_{\vec{A}}$, or even just $\LT$, for 
for the mapping $\pi^{\ast}_{\vec{A}} : \V(A) \rightarrow \tilde{\V}(\vec{A})$,  whenever context makes it convenient and relatively unambiguous to do so.

We have canonical mappings
\[\pi : \Vst(A_1) \times \cdots \times \Vst(A_k) \rightarrow \tilde{\V}(A_1,...,A_k)^{\ast}, 
\ \mbox{and} \ \ m : \V(A_1) \times \cdots \times \V(A_k)   \rightarrow \tilde{\V}(A_1,...,A_k)\]
given by 
\[\pi(a_1,...,a_n)(\mu) = \mu(a_1,...,a_n) \ \ \mbox{and} \ \ m(\alpha_1,...,\alpha_n)(a_1,...,a_n) = \Pi_{i=1}^{n} a_i(\alpha_i).\]
It is straightforward that $(\tilde{\V}(A_1,...,A_n), \pi, m)$ is a composite of $\V(A_1),...,\V(A_n)$. 

For any vector spaces $\V_1,...,\V_k$, let $\L^{k}(\V_1,...,\V_k)$ denote the space of $k$-linear forms 
on $\V_1 \times \cdots \times \V_k$. 
Using the canonical isomorphism $\L^{k}(\Vst_{1},...,\Vst_{k}) \simeq \bigotimes_{i=1}^{k} \V_i$, we can identify $\Blambda(A_1,...,A_k)$, as a vector space, with with $\V(A_1) \otimes \cdots \otimes \V(A_k)$, now ordered by a cone $\tilde{\V}(A_1,....,A_n)_+$ with  
\[ \left (\bigmintensor \V(A_{i}) \right )_{+}  \subseteq \Blambda(A_1,...,A_n)_{+} \subseteq 
\left (\bigmaxtensor \V(A_{i})  \right )_{+}.\]
We also have an injective linear mapping from $\Blambda(A_1,...,A_k) \simeq \bigotimes_i \V(A_i)$ into $\V(A_1,...,A_k)$, extending the map $m$ taking $(\alpha_1,...,\alpha_n) \in \Pi_{i} \V(A_i)$ to $\alpha_1 \otimes \cdots \otimes \alpha_n \in \V(\Pi_{i} A_i)$. However, this mapping is, as a rule, not positive. This will 
be illustrated below in the case of real quantum theory.  
This shows that $\Blambda(A_1,...,A_n)_+$ is generally larger than the 
minimal tensor cone. As we'll also see in the next section, it is generally 
strictly smaller than the maximal tensor cone.
 
{\em Further Notation:} In the case of a bipartite system, we will 
sometimes find it convenient to use 
the tensor-like notation 
$\V(A) \boxtimes \V(B)$ for $\tilde{\V}(A,B)$. Thus, $\V(A) \boxtimes \V(B)$ is 
the vector space $\V(A) \otimes \V(B)$, but ordered by the 
cone $\tilde{\V}(A,B)$ generated by local shadows $\tilde{\omega}$ 
of states $\omega \in \V(AB)$.

The following identifies the effect cone of $\Blambda(A_1,...,A_n)$. We 
omit the straightforward proof.

{\bf Lemma 1:} {\em $\Blambda(A_1,...,A_n)^{\ast}_{+} \simeq \E(\Pi_{i} A_i)_+ \cap 
(\bigotimes_i \E(A_i))$. In the bipartite case: 
\[(\V(A) \boxtimes \V(B))^{\ast}_{+} \simeq \V(AB)^{\ast} \cap (\V(A)^{\ast} \otimes \V(B)^{\ast}).\]}

\tempout{
{\em Proof:} We give the proof for the bipartite case; that for the general case is essentially the same. Assume in what follows that we've identified $\E(A) \otimes \E(B)$ with the corresponding 
subspace of $\E(AB)$. 
Let $f \in (\E(A) \otimes \E(B)) \cap \E(AB)_{+}$. If $\omega \in \V(A \boxtimes B)_{+}$, 
then $\omega(a,b) \geq 0$ for all $(a,b) \in \E(A) \times \E(B)$. Then there exists 
some $\mu \in \V(AB)$ with $\omega(a \otimes b) = \mu(a \otimes b)$ for all 
$a, b \in \E(A) \times \E(B)$. It follows then that $\omega = \mu$ on 
$\E(A) \otimes \E(B)$. Since $f \in \E(AB)_{+}$, $f(\mu) \geq 0$ (and, note, is defined!) 
So $f \in \V(A \boxtimes B)^{\ast}_{+}$. 

Now let $f \in \V(A \boxtimes B)^{\ast}_{+}$. Since $\V(A \boxtimes B) = \V(A) \otimes \V(B)$ as vector spaces, $f \in (\V(A) \otimes \V(B))^{\ast} = \E(A) \otimes \E(B)$. Since 
$f(\omega) \geq 0$ for all $\omega \in \V(A \boxtimes B)_{+}$, if $\omega$ is the 
restriction to an element of $(\E(A) \otimes \E(B))^{\ast}$ of a member $\mu$ of $\V(AB)_{+}$, we have $f(\mu) = f(\omega) \geq 0$. $\Box$  
}


\subsection{Shadows of Processes}

Given a probabilistic theory $(\Cat, \V)$, we would now like to construct 
a locally tomographic ``shadow" theory by applying the $\LT$ construction 
to the objects of $\V(\Cat)$. In order to do this, we first need to decide 
what the morphisms should be in this putative ``shadow" theory.  

In what follows, $A = \Pi \vec{A}$ and $B = \Pi \vec{B}$ are composite systems, with specified decompositions 
$\vec{A} = (A_1,...,A_n)$ and $\vec{B} = (B_1,...,B_k)$.  
The proof of the following is routine.

{\bf Lemma 2:} {\em Let  Let $\Phi :  \V(A) \rightarrow \V(B)$ be a positive linear mapping. The following are equivalent: 
\begin{mlist} 
\item[(a)] $\Phi$ maps $\ker(\LT_{\vec{A}})$ into $\ker(\LT_{\vec{B}})$. 
\item[(b)] If $\omega, \omega' \in \V(A)$ are locally indistinguishable, 
so are $\Phi(\omega), \Phi(\omega')$ in $\V(B)$. 
\item[(c)] There exists a linear mapping $\phi : \bigotimes_i \V(A_i) \rightarrow \bigotimes_j \V(B_j)$ such that $\LT_{\vec{B}} \circ \Phi = \phi \circ \LT_{\vec{A}}$
\end{mlist} 
}

{\bf Definition:} With notation as above, call a positive linear mapping $\Phi : \V(A) \rightarrow \V(B)$ satisfying 
any, hence all, of these conditions {\em locally positive} (with respect to 
the specified decompositions).  The 
linear mapping $\phi$ in part (c) is then uniquely determined. We call 
it the {\em shadow} of $\Phi$, writing $\phi = \LT(\Phi)$. 


As an immediate consequence, we have 

{\bf Lemma 3:} {\em If $\Phi : \V(A) \rightarrow \V(B)$ is locally positive, 
then $\phi = \LT(\Phi)$ is positive as a mapping from $\Blambda(A_1,...,A_m)\rightarrow\Blambda(B_1,...,B_n)$. }

Locally positive maps are reasonably abundant, but do exclude some important 
morphisms in $\RQM$. For instance, if 
$\sigma$ and $\alpha$ are swap and 
associator morphisms in $\Cat$, 
$\V(\sigma)$ is locally 
positive, but $\V(\alpha)$ need not be.

 For another example, if $\alpha$ is a state on $A = A_1 \otimes \cdots \otimes A_n$, then the corresponding mapping $\alpha : \R = \V(I) \rightarrow \V(A)$ given by $\alpha(1) = \alpha$ is 
trivially locally positive (the kernel of $\LT_{I}$ is trivial). On the other hand, there is generally no guarantee that 
an effect $f : \V(A) \rightarrow \R$ will be locally positive. Indeed, if 
$\omega, \omega' \in \V(A)_{+}$ are distinct, there will certainly exist 
some positive linear functional $f$ with $f(\omega) \not = f(\omega')$, 
and by re-scaling if necessary, we can take $f$ to be an effect. If 
$\LT(\omega) = \LT(\omega')$, then $f$ is not locally positive. Thus, 
the passage from $\V(A)$ to $\LT(\V(A))$ not only identifies previously 
distinct states, but also jettisons some effects. 

It follows from part (c) of Lemma 3.3 that if $\Phi : \Pi_{i} A_i \rightarrow \Pi_j B_j$ and $\Psi : \Pi_j B_j  \rightarrow \Pi_k C_k$ are 
locally positive with shadows $\phi = \LT(\Phi)$ and $\psi = \LT(\Psi)$, then 
$\Psi \circ \Phi$ is locally positive with shadow $\psi \circ \phi$. 

{\bf Definition:} 
Let $(\Cat, \V)$ be a probabilistic theory. For objects $\vec{A} = (A_1,...,A_n)$ and $\vec{B} = (B_1,...,B_k) \in \Cat^{\ast}$, call a morphism $\Pi \vec{A} \stackrel{f}{\longrightarrow} \Pi \vec{B}$ {\em local} iff 
$\V(f) : \V(\Pi \vec{A}) \rightarrow \V(\Pi \vec{B})$ is locally positive (relative to the preferred factorizations of $A$ and $B$).  We write $\loCat$ for the category having the same objects as $\Cat^{\ast}$ --- finite lists of non-identity objects in $\Cat$ --- but only local morphisms.

By the remarks above, $\loCat$ is a monoidal 
sub-category of $\Cat^{\ast}$. By construction, the functor $\V \circ \Pi : \Cat^{\ast} \rightarrow \Prob$ descends to a functor $\loCat \rightarrow \Prob$. However, because 
$\Pi$ is not injective on objects, neither will $\V \circ \Pi$ be. To remedy this, we 
make the following 

{\bf Definition:} For all objects $\vec{A}, \vec{B} \in \Cat^{\ast}$ and local morphisms $f : \vec{A} \rightarrow \vec{B}$,  let $\tilde{\V}(\vec{A}) := LT_{\vec{A}}(\V(\Pi \vec{A})$ and $\tilde{\V}(f) := \LT(\V(f))$.

{\bf Lemma 4:} {\em $\tilde{\V}$ is a functor $\loCat \rightarrow \Prob$, and is injective on objects.} 

{\em Proof:} It is straightforward that $\tilde{\V}(f \circ g) = \tilde{\V}(f) \circ \tilde{\V}(g)$ where $f \in \Cat^{\ast}(A,B)$ and $g \in \Cat^{\ast}(B,C)$. That $\tilde{\V}$ is 
injective on objects follows from the fact that 
there is a canonical isomorphism between $\bigotimes \V(A_i)$ and $\L^{k}(\V(A_1)^{\ast},...,\V(A_n)^{\ast})$; the latter is 
literally a space of functions on $\V(A_1)^{\ast} \times \cdots \times \V(A_n)^{\ast}$, from which one can read off 
the spaces $\V(A_1)$, ... $\V(A_n)$. As $\V$ is injective 
on objects, these in turn determine $(A_1,...,A_n)$. 
$\Box$

{\bf Definition:} The {\em locally tomographic shadow} of a monoidal probabilistic theory $(\Cat, \V)$ is the probabilistic theory 
$\LT(\Cat, \V) := (\loCat, \tilde{\V})$. 

By construction, $\LT(\V, \Cat) = (\loCat, \tilde{\V})$ is locally tomographic. 
\tempout{Going forward, it will be important to keep in view the distinction between 
(1) $\loCat$, a sub-category of 
$\Cat^{\ast}$; 
$(\loCat,\tilde{\V})$, a probabilistic theory, and 
$\tilde{\V}(\loCat)$,  
(a sub-category of $\Prob$, under our injectivity-on-objects assumption.) 
}
We have the following picture: there are two functors (probabilistic models) 
associated with $\loCat$: one is given by $\V \circ \Pi$, and the other by 
$\tilde{\V} := \LT(\V \circ \Pi)$, i.e., for each object $\vec{A} =  (A_1,...,A_n)$ in $\loCat$, we have a positive linear mapping $\LT_{\vec{A}} : \V(\Pi(\vec{A})) \rightarrow \tilde{\V}(\vec{A})$.  By construction, 
$\LT$ then defines a natural transformation 
$\V \circ \Pi \Rightarrow \tilde{\V}$.

\newpage

\section{The Shadow of Real Quantum Theory} 

We'll now consider the case of finite-dimensional real quantum theory in some detail, concentrating on the bipartite case.

\subsection{The LT map} 

Suppose $\H$ and $\K$ are two finite-dimensional real Hilbert spaces. 
In what follows, we identify states with the corresponding density operators, 
so that if $W$ is a density operator on $\H \otimes \K$, $\LT(W)$ is the unique operator in $\L_{s}(\H) \otimes \L_{s}(\K)$ satisfying 
$\Tr(\LT(W) a \otimes b) = \Tr(W a \otimes b)$ 
for all $a \in \L_{s}(\H)$ and $b \in \L_{s}(\K)$. 

As discussed in the Introduction, $\L_{s}(\H \otimes \K)$ has a natural orthogonal direct sum decomposition as 
\[\L_{s}(\H \otimes \K) \ = \ \left ( \L_{s}(\H) \otimes \L_{s}(\K) \right ) \oplus \left (\L_{a}(\H) \otimes \L_{a}(\K) \right )\]

Since $\ker(\LT)$ contains $\L_{a}(\H) \otimes \L_{a}(\K)$ and $\Ran(\LT)$ equals $\L_{s}(\H) \otimes \L_{s}(\H)$, we have  $\ker(\LT) = \L_{a}(\H) \otimes \L_{a}(\K)$. 

Let $\Sym : \L(\H) \rightarrow \L_{s}(\H)$ be the symmetrization mapping $\Sym(a) = \tfrac{1}{2}(a + a^{t})$. Note that this 
is the orthogonal projection on $\L(\H)$ (with respect to the trace inner product) with range $\L_{s}(\H)$.  
A straightforward 
computation gives us 

{\bf Lemma 5:} {\em $\LT(W) = (\Sym \otimes \Sym)(W)$ for all $W \in \L(\H)$.}



\tempout{{\em Proof:} Since $\Sym \otimes \Sym$ is self-adjoint w.r.t the trace inner 
product, we have 
\[\Tr((\Sym \otimes \Sym)(W) a \otimes b) 
= \Tr(W (\Sym \otimes \Sym)(a \otimes b)) = \Tr(W a \otimes b)\]
for all $a \in \L_{s}(\H)$ and $b \in \L_{s}(\K)$. $\Box$ 
}

\subsection{The locally tomographic cone}

Using the notation introduced in Section 3, $\L_{s}(\H) \ltensor \L_{s}(\H)$ 
stands for $\L_{s}(\H) \otimes \L_{s}(\H)$ as ordered by the 
{\em locally tomographic cone}  
\[(\L_{s}(\H) \boxtimes \L_{s}(\K))_{+} \ := \ \LT(\L_{s}(\H \otimes \H)_{+}).\] 
Let $(\L_{s}(\H) \otimes \L_{s}(\K))_{+}$ 
stand for the cone of positive operators belonging to $\L_{s}(\H) \otimes \L_{s}(\K)$. 
That is, 
\[(\L_{s}(\H) \otimes \L_{s}(\H))_+ \ := \ (\L_{s}(\H) \otimes \L_{s}(\H)) \cap \L_{s}(\H \otimes \H)_{+}.\]
A priori, we have 
\begin{eqnarray*}
(\L_{s}(\H) \mintensor \L_{s}(\K)_+ & \subseteq & (\L_{s}(\H) \otimes \L_{s}(\K))_{+}\\
&  \subseteq &  
(\L_{s}(\H) \boxtimes \L_{s}(\K))_{+} \subseteq (\L_{s}(\H) \maxtensor \L_{s}(\K))_{+}.\end{eqnarray*}
We'll presently see that if $\H$ and $\K$ are of dimension two or greater, all four of these inclusions are proper. 
\tempout{Returning to the relationship between the cones 
$(\L_{s}(\H) \otimes \L_{s}(\K))_{+}$, $(\L_{s}(\H) \boxtimes \L_{s}(\K))_{+}$, 
and $\L_{s}(\H) \maxtensor \L_{s}(\K)$,  
we will establish that the second sits properly between the other two.}
If $x, y \in \H$, we use the standard operator-theoretic notation $x \odot y$ (rather than $|x \rangle \langle y |$ as in Dirac notation) for the operator on $\H$ given by $(x \odot y)z \ = \ \langle z, y \rangle x$ 
for all $z \in \H$. If $\|x\| = 1$, then $x \odot x = P_{x}$, the rank-one projection onto the span of $x$. 

{\bf Example 1:} Let $\H = \R^{2}$ and let $\{x,y\}$ be any 
orthonormal basis. Let $z$ be the real EPR state 
$z \ = \ \tfrac{1}{\sqrt{2}}( x \otimes y + y \otimes x)$. 
\tempout{Then 
{\small 
\begin{eqnarray*}
z \odot z & = & \tfrac{1}{2} ((x \otimes y + y \otimes x) \odot 
(x \otimes y + y \otimes x)\\
& = & 
\frac{1}{2} (((x \otimes y) \odot (x \otimes y) + (y \otimes x) \odot (y \otimes x)) + ((x \otimes y) \odot (y \otimes x) + (y \otimes x) \odot (x \otimes y)))\\
& = & 
\frac{1}{2} (((x \odot x) \otimes (y \odot y) + (y \odot y) \otimes (x \odot x)) + ((x \odot y) \otimes (y \odot x) + (y \odot x) \otimes (x \odot y)))\\
\end{eqnarray*}
}}
Direct computation 
shows that 
\[z \odot z \ = \ \tfrac{1}{2}(P_x \otimes P_y + P_y \otimes P_x + (x \odot y) \otimes (y \odot x) + (y \odot x) \otimes (x \odot y))\]
Now, $\Sym(x \odot y) = \tfrac{1}{2}(x \odot y + y \odot x) = \Sym(y \odot x) =: S$ 
where $Sx = \tfrac{1}{2}y$ and $Sy = \tfrac{1}{2}x$. So
$W := \LT(z \odot z) \ = \ \tfrac{1}{2}(P_{x} \otimes P_{y} + P_{y} \otimes P_{x})
+ S \otimes S$. This is not a positive operator. For instance, 
if $v =  x \otimes x - y \otimes y$, then $Wv = -\tfrac{1}{4} v$. 

This shows that the embedding $\V(A \boxtimes B) \rightarrow \V(AB)$ is in general not positive, as mentioned earlier. Consequently, 
$(\L_{s} \otimes \L_{s})_{+}$ is strictly smaller than $(\L_{s}(\H) \boxtimes \L_{s}(\K))_+$.

{\bf Theorem 1:} \emph{Let $\dim{\H}$ and $\dim{\K}$ be $3$ or greater.  The cone $(\L_{s}(\H) \otimes \L_{s}(\K))_{+}$ is 
strictly larger than the cone $(\L_{s}(\H) \mintensor \L_{s}(\K))_{+}$, 
and the cone $(\L_{s}(\H) \boxtimes \L_{s}(\K))_{+}$  is strictly smaller than the cone} $(\L_{s}(\H) \maxtensor \L_{s}(\K))_{+}$.

\emph{Proof (sketch):} To simplify notation a bit, let 
$\H = \K$, and write $\L_{s}(\H)$ and $\L_{a}(\H)$ as $\L_{s}$ and $\L_{a}$. 
Lemma 1  tells us that 
$(\L_{s} \boxtimes \L_{s})_{+}^{\ast} \simeq (\L_{s} \otimes \L_{s})_{+}$. 
There exist well-known examples of entangled states 
in $\L_{s} \otimes \L_{s}$, namely those arising from 
unextendible product bases; see \cite{Bennett}.\footnote{We thank Giulio Chiribella for drawing our attention to these.} 
Hence, $(\L_{s}\otimes \L_{s}) \cap \L_{s}(\H \otimes \K)_+$ 
is strictly larger than the minimal tensor cone, which contains 
only unentangled states. Dualizing, we see that 
$(\L_{s}(\H)  \boxtimes \L_{s}(\K))_{+}$ must be strictly 
smaller than the maximal tensor cone. $\Box$ 

{\em Remark:} The geometry of the locally tomographic cone 
$(\L_{s}(\H) \boxtimes \L_{s}(\K))_{+}$ appears to be quite 
interesting. Although the mapping $\LT : \L_{s}(\H \otimes \K) \rightarrow \L_{s}(\H) \boxtimes \L_{s}(\K)$ is in general many-to-one, remarkably, this is not the case for 
pure states: as shown by Lemma 17 of \cite{CDP}, any pure state of $\H \otimes \K$ can be distinguished from any other state (pure or mixed) by product effects. Thus, if $\omega$ is a pure state in $\L_{s}(\H \otimes \K)$, it is the {\em only} state 
with local shadow $\LT(\omega)$.\footnote{If $\omega$ is an interior point in the cone $\L_{s}(\H \otimes \K)_{+}$, then the affine space $\omega + \L_{a}(\H) \otimes \L_{a}(\K)$, whose elements $\mu$ all satisfy $\tr (\mu (a \otimes b)) = \tr (\omega (a \otimes b))$ for all product
effects $a \otimes b$, will certainly intersect the boundary of the positive cone. However, this intersection will not contain pure states, but only points interior to higher-dimensional faces of the cone.}
\tempout{
products of pure states. 

{\bf  Theorem 2:} {\em If $\H$ and $\K$ are finite dimensional real Hilbert spaces and $u \in \H$ and $v \in \K$ are unit vectors, then 
the only density operator  $W \in \L(\H \otimes \K)_{s}$ with $\LT(W) = P_{u} \otimes P_{v}$ is $P_{u} \otimes P_{v}$. } 

For the proof, see the Appendix.
}

\subsection{Processes} 

Let $\Phi : \L_{s}(\H \otimes \K) \rightarrow \L_{s}(\H \otimes \K)$ be a positive linear mapping. For simplicity of notation, let's write $\L_{s,s}$ for $\L_{s}(\H) \otimes \L_{s}(\K)$, 
$\L_{s,s}$ and $\L_{a,s}$ for $\L_{a}(\H) \otimes \L_{a}(\K)$, and so on.
We then have an orthogonal decomposition 
\[\L_{s}(\H \otimes \K) = \L_{s,s} \oplus \L_{s,a} \oplus \L_{a,s} \oplus \L_{a,a},\]
with  respect to which 
$\Phi$ has an operator matrix $\left [ \begin{array}{cc} \Phi_{s,s} & \Phi_{s,a} \\ \Phi_{a,s} & \Phi_{a,a} \end{array} \right ]$, where, 
e.g, $\Phi_{s,s} : \L_{s,s} \rightarrow \L_{s,s}$, $\Phi_{a,s} : \L_{s,s} \rightarrow \L_{a,a}$, etc.  
A straightforward translation of Lemma 2 gives us 

{\bf Lemma 6:} {\em Let $\Phi$ be as above. Then $\Phi$ is locally positive 
iff $\Phi_{s,a} = 0$, 
and in this case, $\LT(\Phi) = \Phi_{s,s}$. } 

This provides another way to see that effects on $\H \otimes \K$, as processes 
\[\L_{s}(\H \otimes \K) \rightarrow \L_{s}(\R),\] 
are generally not locally positive. The following example is particularly noteworthy:

{\bf Example 2:} Consider the functional $\epsilon : \L(\R^2 \otimes \R^2) = \L(\R^2) \otimes \L(\R^2) \rightarrow \R$ corresponding to the trace inner product, i.e., the functional uniquely defined on 
pure tensors by $\epsilon (a \otimes b) = \Tr(a b^{t})$.  
The subspace  
$\L_{a,a} \leq \L(\R^2)$ is one-dimensional, spanned by the operator $J \otimes J = \J$, where $J(x,y) = (-y,x)$. 
Note that $J^2 = -\1$. Hence 
$\epsilon (\J) = \Tr(J J^t) = -\Tr(J^2) = \Tr(\1) = 2$. 
Since $\epsilon$ does not vanish on $\L_{a,a}$, $\epsilon$ is 
not locally positive.

\section{Conclusions and questions} 

At a minimum, $\LT(\RQM)$ provides us with an interesting ``foil"  GPT, 
related to but distinct from both complex and real finite-dimensional real quantum theory, and 
from their Jordan-algebraic relatives \cite{CCEJA} (which, like $\RQM$, 
are not locally tomographic). Among the many questions that suggest themselves about  this theory, and 
about the $\LT$ construction more generally, 
the following stand out to us as particularly interesting. 

{\em Compact Closure} Example 2 shows that 
$\LT(\RQM)$ does not inherit the usual compact structure from $\RQM$. Given a monoidal probabilistic theory, 
theory $(\Cat, \V)$ with $\Cat$ compact closed, when {\em is} $\LT(\Cat, \V)$ 
compact closed?

\tempout{
This raises the question 
of whether $\LT(\RQM)$ is closed for {\em any} choice of duals, units and co-units. More generally, 
More generally, given a probabilistic theory 
$(\Cat, \V)$ with $\Cat$ compact closed, when is its locally tomographic shadow also compact closed? 
}

\tempout{
{\bf Question 2:}{\em Bi-monoidal structure} 
{\em Pure product states} If $\alpha$ and $\beta$ are states of $A$ and $B$, respectively, can there exist a state $\omega \in \Omega(AB)$ with $\omega \not = \alpha \otimes \beta$ but $\LT(\omega) = \alpha \otimes \beta$? If $\alpha$ and $\beta$ are pure, our Theorem 4.6 shows that the answer is {\em no} for composites of real quantum systems, but the 
question is open for more general states and probabilistic theories. 
}

{\em $\LT$ and Complex QM} 
The functor $R : \CQM \rightarrow \RQM$ given 
by restriction of scalars does not preserve tensor products. It would be of interest to understand the functor $\LT \circ R$ from $\CQM$ to $\LT(\RQM)$. 
One can ask a parallel question about complexification.

{\em The Shadow of InvQM}  In \cite{CCEJA}, we constructed a non-locally-tomographic theory we called {\bf InvQM}, which contains finite-dimensional real and quaternionic QM as sub-theories, and also a relative of complex QM in which the composite of two complex quantum systems comes with an extra binary superselection rule. As we will discuss elsewhere, much of what was done above for $\RQM$ generalizes readily to $\InvQM$, but the resulting theory --- like $\LT(\RQM)$ --- remains largely unexplored.  

{\em Non-deterministic shadows} If local agents Alice and Bob agree that their joint state  is $\omega$, this is consistent with the actual, global state being any element  $\mu \in \LT_{A,B}^{-1}(\omega)$. If the (unknown) actual state evolves under a (global) process $\phi : \V(AB) \rightarrow \V(CD)$, the result will be one of the states  in $\phi(\LT_{A,B}^{-1}(\omega))$. Unless $\phi$ is local, these will not be confined to a single fibre of $\LT_{C,D}$; parties $C$ and $D$ might observe any of the different  states in $\LT_{C,D}(\phi(\LT_{A,B}^{-1}(\omega)))$, giving the impression that  $\phi$ acts indeterministically. 
How ought this uncertainty be quantified?  Note that in this situation, the states of $AB$ act as hidden variables ``explaining" this apparent lack of determinism.

\tempout{

{\bf Question 6:} What do $\LT$ and its kernel look like if we treat the real and quaternionic Hilbert spaces $\H$ and $\K$ {\em \`{a} la} Baez, as 
complex spaces with a real or quaternionic structure?

{\bf Question 7:} What is the locally tomographic shadow of complex QM 
as Baezically embedded in $\RQM$ or in $\InvQM$? 

{\bf Question 8:}  A related question: what happens to the locally tomographic shadow of real QM when we complexify? The results of \cite{CDPR} may be relevant here. 


{\bf Question 9:} Let $\RHilb$ be the category of finite-dimensional real Hilbert spaces and linear mappings, and let $\CP(\RHilb)$ be the result of 
applying Selinger's  CP construction to it. Note that both of these are compact closed.  Let $\RQM$ be mixed-state real quantum theory, with CP maps as morphisms. What's the relationship between 
$\CP(\RHilb)$ and $\LT(\RQM)$?
}

\nocite{*}
\bibliographystyle{eptcs}
\bibliography{BGW}

\begin{thebibliography}{10}
\providecommand{\bibitemdeclare}[2]{}
\providecommand{\surnamestart}{}
\providecommand{\surnameend}{}
\providecommand{\urlprefix}{Available at }
\providecommand{\url}[1]{\texttt{#1}}
\providecommand{\href}[2]{\texttt{#2}}
\providecommand{\urlalt}[2]{\href{#1}{#2}}
\providecommand{\doi}[1]{doi:\urlalt{https://doi.org/#1}{#1}}
\providecommand{\eprint}[1]{arXiv:\urlalt{https://arxiv.org/abs/#1}{#1}}
\providecommand{\bibinfo}[2]{#2}

\bibitemdeclare{article}{ALPP}
\bibitem{ALPP}
\bibinfo{author}{Guillaume \surnamestart Aubrun\surnameend},
  \bibinfo{author}{Ludovico \surnamestart Lami\surnameend},
  \bibinfo{author}{Carlos \surnamestart Palazuelos\surnameend} \&
  \bibinfo{author}{Martin \surnamestart Pl{\'a}vala\surnameend}
  (\bibinfo{year}{2021}): \emph{\bibinfo{title}{Entangleability of cones}}.
\newblock {\slshape \bibinfo{journal}{Geometric and Functional Analysis}}
  \bibinfo{volume}{31}(\bibinfo{number}{2}), pp. \bibinfo{pages}{181--205},
  \doi{10.1007/s00039-021-00565-5}.

\bibitemdeclare{article}{CCEJA}
\bibitem{CCEJA}
\bibinfo{author}{Howard \surnamestart Barnum\surnameend},
  \bibinfo{author}{Matthew~A. \surnamestart Graydon\surnameend} \&
  \bibinfo{author}{Alexander \surnamestart Wilce\surnameend}
  (\bibinfo{year}{2020}): \emph{\bibinfo{title}{Composites and categories of
  Euclidean Jordan algebras}}.
\newblock {\slshape \bibinfo{journal}{Quantum}} \bibinfo{volume}{4}, p.
  \bibinfo{pages}{359}, \doi{10.22331/q-2020-11-08-359}.

\bibitemdeclare{article}{BWInf}
\bibitem{BWInf}
\bibinfo{author}{Howard \surnamestart Barnum\surnameend} \&
  \bibinfo{author}{Alexander \surnamestart Wilce\surnameend}
  (\bibinfo{year}{2011}): \emph{\bibinfo{title}{Information Processing in
  Convex Operational Theories}}.
\newblock {\slshape \bibinfo{journal}{Electronic Notes in Theoretical Computer
  Science}} \bibinfo{volume}{270}(\bibinfo{number}{1}), pp.
  \bibinfo{pages}{3--15}, \doi{10.1016/j.entcs.2011.01.002}.
\newblock \bibinfo{note}{Proceedings of the Joint 5th International Workshop on
  Quantum Physics and Logic and 4th Workshop on Developments in Computational
  Models (QPL/DCM 2008)}.

\bibitemdeclare{article}{BW12}
\bibitem{BW12}
\bibinfo{author}{Howard \surnamestart Barnum\surnameend} \&
  \bibinfo{author}{Alexander \surnamestart Wilce\surnameend}
  (\bibinfo{year}{2014}): \emph{\bibinfo{title}{Local tomography and the Jordan
  structure of quantum theory}}.
\newblock {\slshape \bibinfo{journal}{Foundations of Physics}}
  \bibinfo{volume}{44}, pp. \bibinfo{pages}{192--212},
  \doi{10.1007/s10701-014-9777-1}.

\bibitemdeclare{incollection}{BWFoils}
\bibitem{BWFoils}
\bibinfo{author}{Howard \surnamestart Barnum\surnameend} \&
  \bibinfo{author}{Alexander \surnamestart Wilce\surnameend}
  (\bibinfo{year}{2016}): \emph{\bibinfo{title}{Post-Classical Probability
  Theory}}.
\newblock In \bibinfo{editor}{Giulio \surnamestart Chiribella\surnameend} \&
  \bibinfo{editor}{Robert~W. \surnamestart Spekkens\surnameend}, editors:
  {\slshape \bibinfo{booktitle}{Quantum Theory: Informational Foundations and
  Foils}}, {\slshape \bibinfo{series}{Fundamental Theories of Physics}}
  \bibinfo{volume}{181}, \bibinfo{publisher}{Springer},
  \bibinfo{address}{Dordrecht}, pp. \bibinfo{pages}{367--420},
  \doi{10.1007/978-94-017-7303-4_11}.

\bibitemdeclare{article}{Bennett}
\bibitem{Bennett}
\bibinfo{author}{Charles~H. \surnamestart Bennett\surnameend},
  \bibinfo{author}{David~P. \surnamestart DiVincenzo\surnameend},
  \bibinfo{author}{Tal \surnamestart Mor\surnameend}, \bibinfo{author}{Peter~W.
  \surnamestart Shor\surnameend}, \bibinfo{author}{John~A. \surnamestart
  Smolin\surnameend} \& \bibinfo{author}{Barbara~M. \surnamestart
  Terhal\surnameend} (\bibinfo{year}{1999}): \emph{\bibinfo{title}{Unextendible
  product bases and bound entanglement}}.
\newblock {\slshape \bibinfo{journal}{Physical Review Letters}}
  \bibinfo{volume}{82}(\bibinfo{number}{26}), p. \bibinfo{pages}{5385},
  \doi{10.1103/PhysRevLett.82.5385}.

\bibitemdeclare{article}{Chiribella}
\bibitem{Chiribella}
\bibinfo{author}{Giulio \surnamestart Chiribella\surnameend}
  (\bibinfo{year}{2021}): \emph{\bibinfo{title}{Process Tomography in General
  Physical Theories}}.
\newblock {\slshape \bibinfo{journal}{Symmetry}}
  \bibinfo{volume}{13}(\bibinfo{number}{11}), p. \bibinfo{pages}{1985},
  \doi{10.3390/sym13111985}.

\bibitemdeclare{article}{CDP}
\bibitem{CDP}
\bibinfo{author}{Giulio \surnamestart Chiribella\surnameend},
  \bibinfo{author}{Giacomo~Mauro \surnamestart D'Ariano\surnameend} \&
  \bibinfo{author}{Paolo \surnamestart Perinotti\surnameend}
  (\bibinfo{year}{2010}): \emph{\bibinfo{title}{Probabilistic theories with
  purification}}.
\newblock {\slshape \bibinfo{journal}{Phys. Rev. A}} \bibinfo{volume}{81}, p.
  \bibinfo{pages}{062348}, \doi{10.1103/PhysRevA.81.062348}.

\bibitemdeclare{article}{CDPR}
\bibitem{CDPR}
\bibinfo{author}{Giulio \surnamestart Chiribella\surnameend},
  \bibinfo{author}{Kenneth~R. \surnamestart Davidson\surnameend},
  \bibinfo{author}{Vern~I. \surnamestart Paulsen\surnameend} \&
  \bibinfo{author}{Mizanur \surnamestart Rahaman\surnameend}
  (\bibinfo{year}{2023}): \emph{\bibinfo{title}{Positive maps and entanglement
  in real Hilbert spaces}}.
\newblock {\slshape \bibinfo{journal}{Annales Henri Poincar{\'e}}}
  \bibinfo{volume}{2023}, pp. \bibinfo{pages}{1--30},
  \doi{10.1007/s00023-023-01325-x}.

\bibitemdeclare{article}{Hardy}
\bibitem{Hardy}
\bibinfo{author}{Lucien \surnamestart Hardy\surnameend} (\bibinfo{year}{2001}):
  \emph{\bibinfo{title}{Quantum theory from five reasonable axioms}}.
\newblock {\slshape \bibinfo{journal}{arXiv preprint quant-ph/0101012}},
  \doi{10.48550/arXiv.quant-ph/0101012}.

\bibitemdeclare{article}{KRF}
\bibitem{KRF}
\bibinfo{author}{Matthias \surnamestart Kl{\"a}y\surnameend},
  \bibinfo{author}{Charles \surnamestart Randall\surnameend} \&
  \bibinfo{author}{David \surnamestart Foulis\surnameend}
  (\bibinfo{year}{1987}): \emph{\bibinfo{title}{Tensor products and probability
  weights}}.
\newblock {\slshape \bibinfo{journal}{International Journal of Theoretical
  Physics}} \bibinfo{volume}{26}, pp. \bibinfo{pages}{199--219},
  \doi{10.1007/BF00668911}.

\bibitemdeclare{article}{Wilce92}
\bibitem{Wilce92}
\bibinfo{author}{Alexander \surnamestart Wilce\surnameend}
  (\bibinfo{year}{1992}): \emph{\bibinfo{title}{Tensor products in generalized
  measure theory}}.
\newblock {\slshape \bibinfo{journal}{International Journal of Theoretical
  Physics}} \bibinfo{volume}{31}, pp. \bibinfo{pages}{1915--1928},
  \doi{10.1007/BF00671964}.

\bibitemdeclare{article}{wilce2018shortcut}
\bibitem{wilce2018shortcut}
\bibinfo{author}{Alexander \surnamestart Wilce\surnameend}
  (\bibinfo{year}{2018}): \emph{\bibinfo{title}{A Shortcut from Categorical
  Quantum Theory to Convex Operational Theories}}.
\newblock {\slshape \bibinfo{journal}{Electronic Proceedings in Theoretical
  Computer Science}} \bibinfo{volume}{266}, pp. \bibinfo{pages}{222--236},
  \doi{10.4204/eptcs.266.15}.
\newblock \bibinfo{note}{Proc. QPL 2017}.

\bibitemdeclare{incollection}{Wittstock}
\bibitem{Wittstock}
\bibinfo{author}{Gerd \surnamestart Wittstock\surnameend}
  (\bibinfo{year}{1974}): \emph{\bibinfo{title}{Ordered normed tensor
  products}}.
\newblock In \bibinfo{editor}{A.~\surnamestart Hartk{\"a}mper\surnameend} \&
  \bibinfo{editor}{Holger \surnamestart Neumann\surnameend}, editors: {\slshape
  \bibinfo{booktitle}{Foundations of Quantum Mechanics and Ordered Linear
  Spaces: Advanced Study Institute Marburg 1973}}, {\slshape
  \bibinfo{series}{Lecture Notes in Physics}}~\bibinfo{volume}{29},
  \bibinfo{publisher}{Springer}, \bibinfo{address}{Berlin, Heidelberg}, pp.
  \bibinfo{pages}{67--84}, \doi{10.1007/3-540-06725-6_10}.

\end{thebibliography}
\end{document}